\documentclass[12pt]{article}
\usepackage{geometry,enumerate,amsmath,amssymb}
\usepackage{fullpage}
\usepackage{graphicx}
\newcommand{\be}{\begin{equation}}
\newcommand{\ee}{\end{equation}}

\renewcommand{\hat}{\widehat}
\renewcommand{\tilde}{\widetilde}
\renewcommand{\epsilon}{\varepsilon}

\newcommand{\iv}{^{-1}}
\def\alphaa {\frac{\alpha}{2}}
\font\mybb=msbm10 at 11pt

\def\bb#1{\hbox{\mybb#1}}

\def\bR {\bb{R}}

\def\half {\frac{1}{2}}

\newcommand{\news}{\setcounter{equation}{0}\quad}
\def\ben{\begin{equation}}
\def\een{\end{equation}}
\def\bea{\begin{eqnarray}}
\def\eea{\end{eqnarray}}
\begin{document}
\title{
\begin{flushright}\ \vskip -2cm {\small {DAMTP-2012-46}}\end{flushright}
\begin{flushright}\ \vskip -1.5cm {\small {DCPT-12/25}}\end{flushright}
\vskip 10pt
\bf{Platonic hyperbolic monopoles}
\vskip 40pt}
\author{
Nicholas S. Manton$^\star$ and Paul M. Sutcliffe$^\dagger$\\[40pt]
{\em \normalsize $^\star$ 
Department of Applied Mathematics and Theoretical Physics,}\\[0pt] 
{\em \normalsize University of Cambridge, Wilberforce Road, 
Cambridge CB3 0WA, U.K.}\\ 
{\normalsize Email: \quad N.S.Manton@damtp.cam.ac.uk}\\[5pt]
{\em \normalsize $^\dagger$ 
Department of Mathematical Sciences,}\\[0pt] 
{\em \normalsize Durham University, Durham DH1 3LE, U.K.}\\ 
{\normalsize Email: \quad p.m.sutcliffe@durham.ac.uk} 
}
\vskip 10pt
\date{Revised version, April 2013,\\
to appear in Commun. Math. Phys.
}
\maketitle
\vskip 30pt
\begin{abstract}
We construct a number of explicit examples of hyperbolic monopoles,
with various charges and often with some platonic symmetry. 
The fields are obtained from instanton data in $\bR^4$ that 
are invariant under a circle action, and in most cases the monopole charge is 
equal to the instanton charge. A key ingredient is the identification 
of a new set of constraints on ADHM instanton data that are sufficient 
to ensure the circle invariance. Unlike for Euclidean monopoles, 
the formulae for the squared Higgs field magnitude in the examples 
we construct are rational functions of the coordinates. Using these formulae,
we compute and illustrate the energy density of the monopoles. We also
prove, for particular monopoles, that the number of zeros of the 
Higgs field is greater than the monopole charge, confirming numerical results
established earlier for Euclidean monopoles. We also present some 
one-parameter families of monopoles analogous to known scattering 
events for Euclidean monopoles within the geodesic approximation.

\end{abstract}
\newpage
\section{Introduction}\news
The Bogomolny equation for $SU(2)$ BPS monopoles in 
Euclidean space is integrable, but only in a few special cases has it 
been actually integrated to yield explicit monopole solutions. Monopole
solutions have a topological charge $N$, and we refer to them
as $N$-monopoles. The spherically symmetric 1-monopole is explicitly 
known, but for the general 2-monopole \cite{Wa8,FHP} and the axially 
symmetric $N$-monopole \cite{PR} there are explicit formulae for the 
fields only on certain symmetry axes. For more general monopoles, the 
formulae contain parameters subject to transcendental constraints. 

Perhaps the most effective approach to constructing $N$-monopoles
in Euclidean space is the Nahm transform \cite{Nahm}, in which
solutions of the Bogomolny equation are obtained from solutions of 
the Nahm equation, a set of nonlinear ordinary differential equations 
for a triplet of $N\times N$ matrices. For $N>2$ the general solution 
of the Nahm equation is not tractable, as it requires explicit data 
regarding theta functions associated with a complex spectral curve 
of genus $(N-1)^2$ \cite{Hit}, and this data is rather
implicit beyond the elliptic case ($N=2$). However, particular
solutions of the Nahm equation have been obtained \cite{HMM,HS2} that 
give rise to monopoles with some platonic symmetry, that is, symmetry
under one of the special discrete subgroups $K$ of $SO(3)$. Here, 
the quotient of the spectral curve by the platonic symmetry 
is an elliptic curve. Even for these
platonic examples there are no explicit formulae for the monopole fields, 
as the Nahm transform requires a numerical implementation \cite{HS1}.
The numerical results display interesting features regarding the 
distribution of the monopole energy density and the number of zeros of the 
Higgs field \cite{Su4}.

In this paper, it is shown that we can improve on the above by turning
to the hyperbolic setting. We make use of Atiyah's observation 
\cite{At} that hyperbolic monopoles may be identified with 
circle-invariant Yang--Mills instantons, provided that the magnitude of the 
Higgs field at spatial infinity is suitably tuned to the curvature 
of hyperbolic space. Our strategy for the construction of platonic 
hyperbolic monopoles is to restrict to Atiyah's simplest tuned case 
and to identify instantons with the required commuting platonic and 
circle symmetries. There are two different ways to impose the
commuting symmetries, depending upon which symmetry acts most 
naturally, and we describe and implement both methods.

Platonic hyperbolic monopoles are qualitatively similar to the
Euclidean monopoles with platonic symmetry. However, many are 
expected to have fields that are rational functions of the
coordinates, and finding these is the main goal of this paper. For 
the spherically-symmetric hyperbolic 1-monopole, in the tuned cases, 
a direct calculation confirms that the Higgs field magnitude is
rational, and taking the flat space limit reveals why the Euclidean 
1-monopole is not rational. Study of hyperbolic monopoles is also 
motivated by their connection with monopoles in Anti-de Sitter 
spacetime \cite{Su3}, and by their likely connection with Skyrmions 
of minimal energy \cite{MSam,AS}.

There is easy access to a large class of instantons which are rational. 
These are the JNR instantons \cite{JNR}, constructed using a formula first 
investigated by Corrigan and Fairlie \cite{CF}. Some of these 
have the circle invariance required for obtaining 
hyperbolic monopoles. We identify a subset of JNR instantons, and the
corresponding monopoles, that also have platonic symmetry.   

We know from previous studies that there are further instantons with 
platonic symmetry. These are obtained using the ADHM formalism 
\cite {ADHM}, from which one can obtain all instantons. There are 
quadratic constraints on the quaternionic ADHM matrices, which in 
general cannot be solved explicitly. So the general instanton is
not rational. However, if the instanton has platonic symmetry and
suitably small charge, 
then the ADHM constraints simplify and can be explicitly solved.
We have discovered that many of these platonic ADHM instantons are 
simultaneously invariant under a commuting circle action. This had 
not been previously realised. We can therefore construct platonic
hyperbolic monopoles from these instantons, and their fields are
rational. In particular, from the squared Higgs field magnitude 
the energy density can be computed by differentiation, so this 
is also rational. 

It is well known that for monopoles rather generally, the number of
zeros of the Higgs field, counted with multiplicity, equals the
topological charge. Well-separated
single monopoles have one Higgs zero each, so here the number of Higgs 
zeros equals the charge. But for more compact monopoles of higher
charge, including those with platonic symmetry, there can be more
zeros than the charge. 
This is a surprising result, given that a theorem of
Jaffe and Taubes \cite{JT} rules out this
possibility for the analogous situation of abelian Higgs vortices
in the plane.
For $N>2$ it often happens that there are $N+n$ zeros 
of multiplicity $1$, for some positive $n$, and $n$ zeros of multiplicity 
$-1$, which are termed anti-zeros \cite{Su4}. A simple example is the
tetrahedrally-symmetric, Euclidean monopole of charge 3. Here the 
symmetry suggests that there are four zeros of positive multiplicity 
at the vertices of a tetrahedron, and one zero of negative
multiplicity at the centre. This has been confirmed by numerical
calculation. Zeros are robust, so there are the same five zeros for
monopoles close to the tetrahedral monopole in the 3-monopole moduli
space. For our platonic hyperbolic monopoles, we are able to calculate the 
locations of Higgs zeros explicitly, using the rational formulae for 
the Higgs field. We find the same arrangement of zeros in the
hyperbolic monopoles as in their Euclidean counterparts.

In Section 2 we introduce our notation and
review some details of hyperbolic monopoles. 
With this in hand, we are then able to present 
a more detailed outline of this paper and set our results within
the context of previous studies.
 
\section{Hyperbolic monopoles}\news
Hyperbolic monopoles \cite{At,Cha,Na} are solutions 
of the Bogomolny equation
\be
*F = D\Phi \,.
\label{Bogo}
\ee
Here $F$ is the field strength of an $SU(2)$ gauge
potential $A$, and $D\Phi$ is the covariant derivative of an adjoint Higgs
field $\Phi$. The hyperbolic geometry enters through the Hodge star,
$*$. The boundary condition is that 
the magnitude of the Higgs field, $|\Phi|,$ has a fixed positive value $v$
at infinity. 
Here $|\Phi|^2=-\frac{1}{2}\mbox{Tr}(\Phi^2).$ 

We will work on the hyperbolic space $\bb{H}^3$ of fixed sectional curvature
$-1$. It will be most convenient to represent $\bb{H}^3$ by the
unit ball model, where the metric is
\be
ds^2(\bb{H}^3)=\frac{4(dX_1^2+dX_2^2+dX_3^2)}{(1-R^2)^2} \,,
\label{metricball0}
\ee
with $R^2=X_1^2+X_2^2+X_3^2$ and $R<1$. In these coordinates the
metric is rational, and monopoles may be rational too. 
In terms of standard spherical polars, given by the relations
$X_1 = R \sin\theta \cos\phi$, 
$X_2 = R \sin\theta \sin\phi$, $X_3 = R \cos\theta$, the metric becomes
\be
ds^2(\bb{H}^3)=\frac{4(dR^2 + R^2(d\theta^2 + \sin^2\theta \, d\phi^2))}
{(1-R^2)^2} \,.
\label{metricballpolar}
\ee 
The geodesic distance from the origin, $\rho$, is related to the radius 
$R$ by $R = \tanh(\rho/2)$. Using $\rho$ as 
radial coordinate, the metric (\ref{metricballpolar}) becomes
\be
ds^2(\bb{H}^3)=d\rho^2 
+ \sinh^2 \rho\,(d\theta^2 + \sin^2\theta \, d\phi^2) \,.
\label{metricgeod}
\ee
The final description of $\bb{H}^3$ that we will need is  
the upper half space model, 
\be ds^2(\bb{H}^3) = \frac{1}{r^2}(dx_1^2 + dx_2^2 + dr^2) \,, 
\label{uhsmetric}\ee
with coordinates $x_1,x_2,r$, 
where $r>0$. 
The relations between the upper half space coordinates and the coordinates 
$X_1,X_2,X_3$ in the unit ball model are
\be
r=\frac{1-R^2}{1+R^2-2X_3} \,, \quad\quad
x_1+ix_2=\frac{2(X_1+iX_2)}{1+R^2-2X_3} \,.
\label{ball2uhs}
\ee 
 
The Bogomolny equation for monopoles in flat space is also
(\ref{Bogo}), but with the Hodge star of Euclidean $\bR^3$ 
\cite{AH,book,Shn}. In flat space, the boundary value $v$ sets
the (inverse) length scale, and replacing $v$ by $\tilde v > v$ just results
in monopoles being scaled down by a factor $\tilde v/v$. Hyperbolic 
space, on the other hand, has a built-in length scale, and the value 
of $v$ affects the monopole solutions in a non-trivial way. 

Hyperbolic monopoles exist for all positive $v$, but
only if $2v$ is an integer, denoted by $2p$, can a 
hyperbolic monopole be interpreted as an $SU(2)$ Yang--Mills 
instanton in $\bR^4$ invariant under a circle action. 
Recall that instantons are solutions of the conformally invariant 
self-dual Yang--Mills equation in $\bR^4$, $*F = F$.  
The hyperbolic monopoles we will consider are circle-invariant 
instantons, and most of them are additionally symmetric under 
some subgroup $K$ of $SO(3)$. We therefore need to review the 
geometry of the commuting $SO(2)$ and $SO(3)$ actions and the
relation between circle-invariant instantons and hyperbolic monopoles.

$SO(3) \times SO(2)$ acts isometrically, as a subgroup of $SO(5)$, 
on Euclidean $\bR^5$. Since this action preserves lengths,
it can be restricted to the unit 4-sphere, $S^4$. The generic orbits of
the group on $S^4$ are $S^2 \times S^1$, and there is a one-parameter
family of these, with the parameter lying in an open interval. At the
ends of the interval are two special orbits. At one end, $S^2$
collapses to a point and the orbit is $S^1$; at the other end $S^1$
collapses to a point and the orbit is $S^2$.

$\bb{H}^3$ is obtained by quotienting by the $SO(2)$ action. To avoid
singularities, $SO(2)$ has to act freely, so the special $S^2$ orbit needs to
be removed and then
\be
S^4 - S^2 \equiv \bb{H}^3 \times S^1 \,.
\ee
This is, in fact, a conformal equivalence. The standard metric on 
$\bb{H}^3 \times S^1$ is conformal to the standard round metric on 
$S^4 - S^2$. The curvature of $\bb{H}^3$ is correlated with the length
of the circle. If we normalise the length of the circle to be $2\pi$, then   
$\bb{H}^3$ has curvature $-1$.

The way to see this is to represent $S^4$ conformally as
Euclidean $\bR^4$ (compactified by a point at infinity). Since the
self-dual Yang--Mills equation is conformally invariant,
instantons on $S^4$ are equivalent to instantons on $\bR^4$ with
appropriate boundary conditions. The latter setting for instantons 
is easier to implement. Let $\bR^4$ have Cartesian coordinates $x_\mu$ 
$(\mu=1,\ldots,4)$ and metric
\be
ds^2 = dx_1^2 + dx_2^2 + dx_3^2 + dx_4^2 \,.
\label{EuclR4}
\ee
Now let $x_3+ix_4=re^{i\chi}$, so $r \ge 0$ and the range of $\chi$ is
$2\pi$. We can define a circle action on $\bR^4$
by the standard rotation of $\chi$. Its fixed point set is the plane $x_3 =
x_4 = 0$, which extends to a 2-sphere in the compactification. We
remove this plane from $\bR^4$ and quotient by the circle action. This
gives $\bb{H}^3$.

Metrically, we re-express (\ref{EuclR4}) as 
\be
ds^2 = dx_1^2 + dx_2^2 + dr^2 + r^2 d\chi^2 \,,
\ee
and note that for $r>0$ this is conformally equivalent to
\be
ds^2 = \frac{1}{r^2}(dx_1^2 + dx_2^2 + dr^2) + d\chi^2 \,,
\ee
which is the product metric on $\bb{H}^3 \times S^1$.
Quotienting by $SO(2)$ gives the metric (\ref{uhsmetric}) 
on $\bb{H}^3$ in the upper half space model. Note that the 
removed plane (plus the point at infinity) can be interpreted as the 
boundary of $\bb{H}^3$.

The isometry group of $\bb{H}^3$ is 6-dimensional, and has no
canonical $SO(3)$ subgroup. However, if we choose a particular point as
the origin of $\bb{H}^3$, then there is a unique $SO(3)$ isometry group
with this as fixed point. We select as origin the point with
coordinates $x_1=x_2=0$ and $r=1$. The orbits of the $SO(3)$ action are then
the 2-spheres $x_1^2 + x_2^2 + r^2 - 2\nu r + 1 = 0$, with $\nu \ge 1$.

The ball model of $\bb{H}^3$ arises
from a different, but conformally equivalent, quotient of $\bR^4$ by a 
circle action. The $SO(2)$ action is slightly more complicated, but the 
$SO(3)$ action is simpler. We introduce toroidal coordinates
$(\rho,\theta,\phi,\chi)$ on $\bR^4$ via
\be
x_\mu = \frac{1}{\cosh\rho + \cos\chi}
(\sinh\rho \sin\theta \cos\phi \,,  \sinh\rho \sin\theta \sin\phi \,,
\sinh\rho \cos\theta \,, \sin\chi) \,.
\ee
Then the flat metric (\ref{EuclR4}) becomes
\be
ds^2 = \frac{ds^2(\bb{H}^3) + d\chi^2}{(\cosh\rho + \cos\chi)^2} \,,
\label{confmetric}
\ee
with $ds^2(\bb{H}^3)$ given by (\ref{metricgeod}).
The flat metric (\ref{confmetric}) is clearly
conformally equivalent to
\be
ds^2 = ds^2(\bb{H}^3) + d\chi^2 \,
\ee
and the quotient by $SO(2)$ is therefore the metric on $\bb{H}^3$,
in the form of the hyperbolic ball model (\ref{metricgeod}),
with $0 \le \rho < \infty$ the geodesic distance from the origin. 
The $SO(3)$ orbits are the 2-spheres
of constant $\rho$, with $\theta$ and $\phi$ usual polar coordinates.

In terms of the earlier unit ball model, with Cartesian coordinates 
$X_1,X_2,X_3,$ this Cartesian ball can be identified with the unit ball in the
hyperplane $x_4=0$ of $\bR^4$, centred at the origin. Each circle
(parametrised by $\chi$) intersects this once, so we may also regard 
$X_1, X_2, X_3$ and $\chi$ as toroidal coordinates on $\bR^4$. 
The special $SO(3)$ orbit, where
the circles collapse to points, can again be identified as the
boundary of $\bb{H}^3$, which is now the 2-sphere, $R=1$. 

We have seen that the quotient of $\bR^4$ by the circle action is 
conformally $\bb{H}^3$, so a circle-invariant gauge potential in 
$\bR^4$ gives rise to a gauge potential on $\bb{H}^3$ together with 
an adjoint Higgs field (the component of the gauge potential along the 
circles), by the standard ideas of dimensional reduction \cite{book}. 
The self-dual Yang--Mills equation reduces to
the Bogomolny equation on $\bb{H}^3$. Atiyah showed that the instanton
charge $I$ and monopole charge $N$ are related by \cite{At} 
\be
I = 2pN \,.
\ee
The boundary value $p$ arises from the way the circle action lifts to
the bundle carrying the $SU(2)$ instanton over the fixed $S^2$ of
$\bR^4$ under the circle action. As discussed above, this $S^2$
is the boundary of $\bb{H}^3$. The simplest case is $p=\half$.
For this value of $p$ the monopole charge and instanton
charge are equal. 

The spherically-symmetric, hyperbolic 1-monopole is explicitly known 
for all $v$. In terms of the coordinates (\ref{metricgeod}), 
the Higgs field has magnitude \cite{Cha,Na}
\be
|\Phi| = \frac{C}{2} \coth C\rho - \frac{1}{2} \coth \rho
\label{SpherHiggs}
\ee
with asymptotic value $v = \half(C-1)$. $C$ takes any value greater 
than 1. Note that $|\Phi|$ varies linearly with $\rho$ near $\rho=0$,
as the pole terms in $\rho$ cancel. This 1-monopole arises from a 
circularly symmetric instanton if and only if $C$ is integral, in 
which case $p$ ($=v$) is half-integral. As shown by the following 
short calculation, for such values of $C$, $|\Phi|$ is a rational 
function of $R$, the radial coordinate in the rational metric 
(\ref{metricball0}).

We rewrite $|\Phi|$ as
\be
|\Phi| = \frac{C(e^{C\rho} + e^{-C\rho})}{2(e^{C\rho} - e^{-C\rho})}
- \frac{e^{\rho} + e^{-\rho}}{2(e^{\rho} - e^{-\rho})}
\label{Phiexp}
\ee
and note that $R = \tanh(\rho/2)$ implies that $e^\rho = \frac{1+R}{1-R}$.
For integer $C$, expression (\ref{Phiexp}) is then clearly a rational
function of $R$. For the first few values of $C$, and the corresponding
$p = \half(C-1)$, this yields
\bea
|\Phi| &=& \frac{R}{1+R^2}\,, \hskip 3.85cm C=2, \ p=\half \\
|\Phi| &=& \frac{8R(1+R^2)}{(3+R^2)(1+3R^2)}\,, \hskip 1.8cm C=3, \ p=1 
\label{pone}\\
|\Phi| &=& \frac{R(5+14R^2+5R^4)}{(1+R^2)(1+6R^2+R^4)}\,, \quad\quad
 C=4, \ p=\frac{3}{2} \,.
\eea
The linear behaviour of $|\Phi|$ near $R=0$, and the asymptotic 
value, $|\Phi| = p$ at $R=1$, are both easily verified.

We can get some insight into the difference between hyperbolic and
Euclidean monopoles by rederiving the Euclidean formula for
$|\Phi|$. For this we need the expression \cite{Cha,Na} for $|\Phi|$ in
hyperbolic space of curvature $-\kappa^2$,
\be
|\Phi| = \frac{C\kappa}{2} \coth (C\kappa \rho) 
- \frac{\kappa}{2}\coth (\kappa \rho) \,.
\label{SpherHiggskappa}
\ee
The Euclidean monopole with $v=1$ is obtained by taking the limit 
$\kappa \to 0$ and $C \to \infty$, with $C\kappa$ fixed to be 2. The 
result is \cite{PS,Bo}
\be
|\Phi| = \coth 2\rho - \frac{1}{2\rho} \,,
\label{SpherHiggsEucl}
\ee
where $\rho$ is the usual radial coordinate in $\bR^3$. This familiar but
rather peculiar expression is rational neither as a function of
$\rho$ nor as a function of $e^\rho$. This is because it arises
from the limit $C \to \infty$. It is not surprising that Euclidean
monopoles of higher charge are not rational either.   

We will discuss circle-invariant instantons and the corresponding 
hyperbolic monopoles in some generality. For most of these, the 
boundary Higgs field will have magnitude $p=\half$. We will focus on examples 
that have an additional invariance under a platonic symmetry 
group, $K \subset SO(3)$, the symmetry being clearest in the
hyperbolic ball model. Instantons with 
platonic symmetry have been studied before \cite{LM,SiSu}. There are 
examples with charge 4 and cubic symmetry, and charge 7 with
icosahedral symmetry. What we need to do here is to find 
which of them have an additional commuting circle invariance. This is 
mainly a matter of determining the correct scale size.

We start with the JNR construction \cite{JNR} in Section 3, as it is 
simpler than the general ADHM construction \cite{ADHM}.
The JNR ansatz gives the gauge potential of an instanton in terms of
derivatives of a scalar potential function $\zeta$ in $\bR^4$. $\zeta$ has
singularities, called ``poles'', at $N+1$ points when the instanton has
charge $N$. The coefficients of the singular terms are called
``weights''. The poles are not singularities of the instanton itself. 
If these poles lie on a plane, $\bR^2$, then $\zeta$ is invariant
under the circle action whose fixed-point set is this plane. This 
leads straightforwardly to a class of hyperbolic monopoles 
defined in the upper half space model of $\bb{H}^3$. To investigate
whether such a monopole has platonic symmetry, we exploit the
conformal invariance of the JNR construction to convert to the ball
model of $\bb{H}^3$. After the conversion, the JNR potential $\zeta$ 
has its poles on the $S^2$ boundary of $\bb{H}^3$, and it is easier to
determine which symmetry group $K$ is present. We also show
how to compute the Higgs field magnitude and energy density. This JNR approach 
gives, in particular, the 3-monopole with tetrahedral symmetry,
analogous to the 3-monopole with the same symmetry in $\bR^3$.

We next discuss circle-invariant ADHM data. A mechanism for
imposing circle invariance and obtaining hyperbolic monopoles was 
established by Braam and Austin \cite{BA}. Their formalism 
applies to the situation where the circle symmetry acts naturally 
in the $(x_3,x_4)$-plane, and is therefore best adapted to hyperbolic 
monopoles in the upper half space model of $\bb{H}^3$, where platonic 
symmetries are not straightforwardly realised. Their analysis works for 
any $p$, and converts the single quaternionic ADHM matrix equation into
a set of coupled complex matrix equations defined on a linear lattice
with $2p$ sites. As mentioned earlier, the Euclidean limit emerges
as $p\rightarrow \infty.$ This is the continuum limit of the lattice system,
and the complex matrix equations turn into the Nahm equation 
for Euclidean monopoles \cite{Nahm}. The lattice system may therefore
be viewed as a discrete Nahm equation \cite{BA}.

The simplest case of the discrete Nahm equation 
is when $p=\half$, where the lattice degenerates to
a single site, and the resulting complex equation is merely the 
original ADHM equation with the quaternionic entries of the ADHM matrix 
restricted to be complex. Braam and Austin did not explicitly
discuss this case, so they did not construct any examples of 
hyperbolic monopoles with $p=\half$, with or without platonic symmetries. 
It is known how to relate the JNR ansatz to a subset of solutions of the 
ADHM equation and our condition that the poles lie in a plane provides the
required restriction from quaternionic to complex data. JNR data 
restricted to a plane therefore provides a subset of solutions to the 
discrete Nahm equation in the degenerate case of one lattice site.

In contrast to the approach of Braam and Austin, our
analysis of circle-invariant ADHM data is based on the ball 
model of $\bb{H}^3$, so that there is a natural action of
$SO(3).$ This means that the circle action is more complicated
than in previous studies of ADHM data. In Section 4 we 
introduce a novel version of circle-invariant ADHM data,
leading to instantons invariant under the circle action on $\bR^4$ 
whose quotient manifestly gives the hyperbolic ball. The associated
hyperbolic monopoles have $p = \half$. The advantage of 
this approach is that several examples of ADHM instanton data with platonic 
symmetry group $K$ have been constructed previously, using 
a systematic approach involving representations of $K$. We have found, perhaps
surprisingly, that many of these examples also satisfy our new 
constraints required for circle invariance, provided the instanton 
scale size is fixed appropriately. ADHM data that simultaneously have 
the circle invariance and platonic symmetry are presented in Section
5. The Higgs field and energy density of the associated
hyperbolic monopoles can be computed explictly with the assistance 
of MAPLE to perform the quaternionic linear algebra. The resulting 
formulae are rational in the unit ball coordinates. 

Although our method yields explicit solutions,
our analysis is less general than that of Braam and Austin.
In particular we have not pinned down the rational map associated with a
general hyperbolic monopole. This is a map that describes the
asymptotic structure of the monopole on the ball boundary, and is
known to completely determine the monopole \cite{BA}. 
We have not yet understood how this rational map arises for our
version of the ADHM data and constraints. However, in Section 6 we 
propose a formula for a rational map that works well for a certain 
class of hyperbolic monopoles. This map is of the Jarvis type 
\cite{Jar}, first defined for monopoles in $\bR^3$, and 
is compatible with the $SO(3)$ action on the hyperbolic ball. 
We do not address spectral curves associated with hyperbolic 
monopoles \cite{MuSi,NoRo}. 

In Section 7 we briefly discuss spherically symmetric hyperbolic
monopoles for other half-integer $p$. Using the upper half space model 
of $\bb{H}^3$, that derives from the planar circle action, we recall 
the JNR version of the required instantons given by Nash \cite{Na}.
We then present the corresponding ADHM data for the unit ball model of
$\bb{H}^3,$ obtained from the more complicated circle action.
This data is then assessed in the light of our new $p=\half$ constraints.

In Section 8 we present our conclusions.

\section{Platonic hyperbolic monopoles via JNR}\news
The JNR ansatz is \cite{JNR}
\be
A_\mu=\frac{i}{2} \, \sigma_{\mu\nu} \, \partial_\nu\log\zeta \,,
\label{jnr}
\ee
where $\sigma_{i4}=\tau_i, \ \sigma_{ij}=\varepsilon_{ijk}\tau_k$,
and $\tau_i \ (i=1,2,3)$ are the Pauli matrices.

Let $\xi_m$, for $m=0,\ldots,N,$ be complex constants and take $\zeta$ to 
have the form
\be
\zeta=\sum_{m=0}^N \frac{1+|\xi_m|^2}{|x_1+ix_2-\xi_m|^2+r^2} \,,
\label{jnrsum}
\ee
where $x_3 + ix_4 = re^{i\chi}$, and the circle action rotates $\chi$.
This gives an $N$-instanton. The $N+1$ singularities of $\zeta$, the 
poles, are all on the fixed plane of the circle action, $r=0$, 
which corresponds to the boundary of $\bb{H}^3$ in the half space 
model. This ensures that the instanton is invariant under the 
circle action, and hence produces a hyperbolic monopole. The poles are
located at the points with complex coordinates $\xi_m$ in this plane. 
The weights of $\zeta$, that is, the numerator factors $1+|\xi_m|^2$, 
have been chosen so that they are all equal
after a conformal transformation to the unit ball model of
$\bb{H}^3$. This is verified using the scaling rule for the weights
under conformal transformations, pointed out in \cite{JNR}. After the
transformation, the poles are on the boundary 2-sphere of 
the hyperbolic ball, and since the weights are equal, the hyperbolic 
monopole acquires the symmetry of the configuration of poles. The
location of the $m$-th pole on the 2-sphere is still $\xi_m$, which
is now the complex coordinate obtained by stereographic projection from
the Cartesian coordinates on the unit sphere by the usual formula
$\xi =(X_1+iX_2)/(1-X_3)$, following from (\ref{ball2uhs}).

The symmetry of the hyperbolic monopole is platonic, if, for example, 
the poles are at the vertices of a platonic solid. The points 
$\xi_m$ then need to be the roots of the vertex Klein polynomial of
that solid \cite{Kle,book}.

As $\partial_\chi\zeta=0$, the only $\chi$-dependence in $A_\mu$
arises from the $\chi$-dependence of the Cartesian partial derivatives in 
the JNR ansatz (\ref{jnr}).
This dependence is removed by the gauge transformation
\be
A_\mu\mapsto GA_\mu G^{-1}-\partial_\mu G\,G^{-1}, \quad
\mbox{where} \quad G=e^{ip\chi\tau_3} \quad \mbox{with} \ \
p=\frac{1}{2} \,.
\ee
In this gauge the Higgs field of the monopole is given by $\Phi=A_\chi$. 
Its magnitude on the boundary of $\bb{H}^3$ is fixed as
$|\Phi|^2 (=-\frac{1}{2}\mbox{Tr}(\Phi^2))$ is equal to $p^2=\frac{1}{4}$
there. As $p = \half$, the instanton number $N$ equals the monopole charge.
Below, we will present formulae only for $|\Phi|$, but $\Phi$ itself
and the gauge potential can easily be found, if required.

The monopole energy density ${\cal E}$ can be written as the 
Laplace--Beltrami operator acting on the squared magnitude of the
Higgs field,
\be
{\cal E}=\frac{1}{\sqrt{g}}\partial_i(\sqrt{g}g^{ij}\partial_j
|\Phi|^2) \,,
\ee
in which the metric $g$ is taken to be the ball metric (\ref{metricball0}).  
This simple expression for the energy density was first derived in
flat space \cite{Wa1}, using the Bogomolny equation, but it easily generalises
to any curved background. The total energy is 
\be
E=\int_{\bb{H}^3} {\cal E}\sqrt{g}\,dX_1dX_2dX_3
=4\pi p N=2\pi N \,.
\ee

\subsection{Spherical 1-monopole}
\begin{figure}[ht]\begin{center}
\includegraphics[width=11cm]{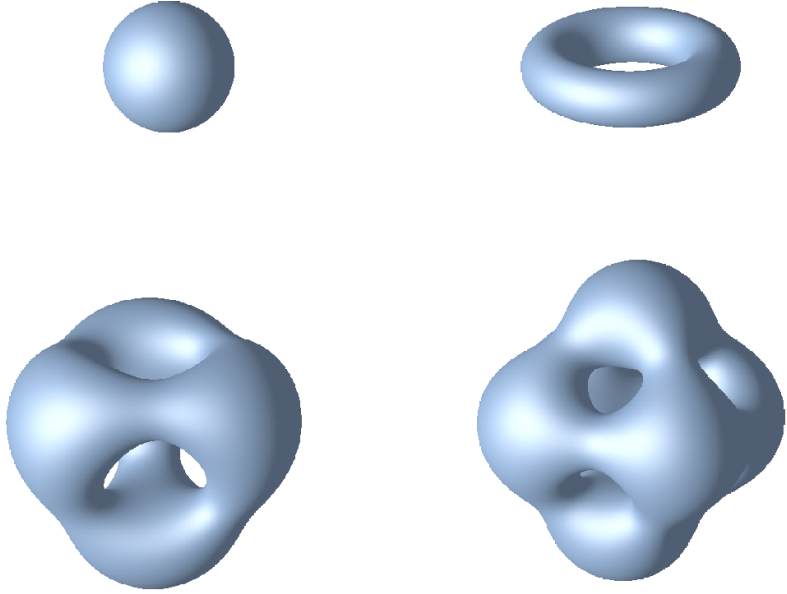}
\caption{Energy density isosurfaces for the spherical 1-monopole,
axial 2-monopole, tetrahedral 3-monopole and octahedral
5-monopole. These surfaces are shown in the ball model of $\bb{H}^3$.}
\label{fig-1235}\end{center}\end{figure}

Taking $\xi_0=-\xi_1=1$ gives a charge 1 hyperbolic monopole. The poles are
antipodal on the 2-sphere, so $\zeta$ is not manifestly spherically
symmetric. However, as observed in \cite{JNR}, this is a case where the 
poles can be moved along any great circle passing through them to another 
antipodal pair of points, just producing a gauge transformation. So 
the monopole is spherically symmetric about the centre of the 
hyperbolic ball. Applying the above formulae yields 
\be
|\Phi|^2=\frac{R^2}{(1+R^2)^2} \,,
\label{1-Higgs}
\ee
with an associated energy density
\be
{\cal E}=\frac{3}{2}\bigg(
\frac{1-R^2}{1+R^2}\bigg)^4 \,.
\label{1-energy}
\ee
Clearly, as $R \to 1$, $|\Phi|^2 \to \frac{1}{4}$ so $p = \half$.
One sees that this basic monopole in hyperbolic space is
indeed rational, making it simpler than 
the flat space monopole, whose Higgs field depends on radius through 
a combination of rational and hyperbolic functions.  
An energy density isosurface is shown in Figure~\ref{fig-1235}.

\subsection{Axial 2-monopole}
Taking $\xi_m=e^{2\pi i m/3}$, with $m=0,1,2$ and 
writing $\rho^2=X_1^2+X_2^2$ yields
\be
|\Phi|^2=
{\frac {{R}^{2} \left( 1+{R}^{2} \right) ^{2}-{\rho}^{2} \left( 1+{R}^
{4} \right) +\frac{1}{4}\,{\rho}^{4}}{ \left(  \left( 1+{R}^{2} \right) ^{2}-{
\rho}^{2} \right) ^{2}}} \,,
\label{axial2jnr}
\ee
with energy density
\be
{\cal E}=
{\frac { \left( 1-{R}^{2} \right) ^{4} \bigl(  \left( 1+{R}^{2}
 \right) ^{4}+22{\rho}^{2} \left( 1+{R}^{2} \right) ^{2}+
4{\rho}^{4} \bigr)}{2 \left(  \left( 1+{R}^{2} \right) ^{2
}-{\rho}^{2} \right) ^{4}}} \,.
\ee

The poles in this case are on the equator of the 2-sphere, located 
at the vertices of an equilateral triangle. This is a triangle that 
can be rigidly rotated around the equator, producing only a gauge 
transformation, so the JNR instanton and the hyperbolic monopole to 
which it gives rise have axial symmetry. An energy density isosurface 
is shown in Figure~\ref{fig-1235}.


\subsection{Tetrahedral 3-monopole}
The vertex Klein polynomial of the tetrahedron is
${\cal T}_v=\xi^4+2\sqrt{3}i\xi^2+1$, with the four roots
$\pm(1+i)/(\sqrt{3}+1), \ 
\pm(1-i)/(\sqrt{3}-1)$. Using the JNR ansatz with these points
$\xi_m$ as poles gives a tetrahedrally symmetric monopole of charge 3. 

The Higgs field and energy density are best expressed, as before, in
terms of the Cartesian coordinates $X_1, X_2, X_3$. 
The ring of tetrahedrally invariant homogeneous polynomials is
generated by the polynomials of degrees two, three and four,
\be
t_2= X_1^2+X_2^2+X_3^2 \,, \quad t_3= X_1X_2X_3 \,, \quad t_4=
X_1^4+X_2^4+X_3^4 \,.
\label{tetrapoly}
\ee
The squared magnitude of the Higgs field can be expressed in terms of 
these. Explicitly, it is found that 
\bea
& & |\Phi|^2=q_1/q_2 \quad \mbox{where} \nonumber\\
&q_1&=9t_2+216\sqrt{3}t_3+132t_2^2-24t_4-24\sqrt{3}t_3t_2+294t_2^3
+48t_4t_2+192t_3^2-24\sqrt{3}t_3t_2^2 \nonumber\\
& &+132t_2^4-24t_4t_2^2+216\sqrt{3}t_3t_2^3+9t_2^5 \,,\nonumber\\
&q_2&=(9+15t_2+16\sqrt{3}t_3+15t_2^2+9t_2^3)^2 \,.
\eea

The zeros of the Higgs field have tetrahedral symmetry, and for the
tetrahedrally symmetric 3-monopole in Euclidean space, it was found (partly
numerically) that there are four zeros of multiplicity $1$ forming a 
tetrahedron, together with one anti-zero at the centre \cite{HS1}. The same 
occurs for the hyperbolic monopole.
Along the line $X_1=X_2=X_3=l/\sqrt{3}$, which passes through a vertex
of the tetrahedron of poles, the above expression simplifies to 
\be |\Phi|^2=
{\frac {{l}^{2} \left( 3{l}^{2}+14l+3 \right) ^{2}}{ \left( 3{l}
^{2}+4l+3 \right) ^{2} \left( 3{l}^{2}-2l+3 \right) ^{2}}} \,,
\ee
which has zeros at $l=0$ and $l=(2\sqrt{10}-7)/3$,
confirming the extra anti-zero of the Higgs field.

\begin{figure}[ht]\begin{center}\includegraphics[width=10cm]{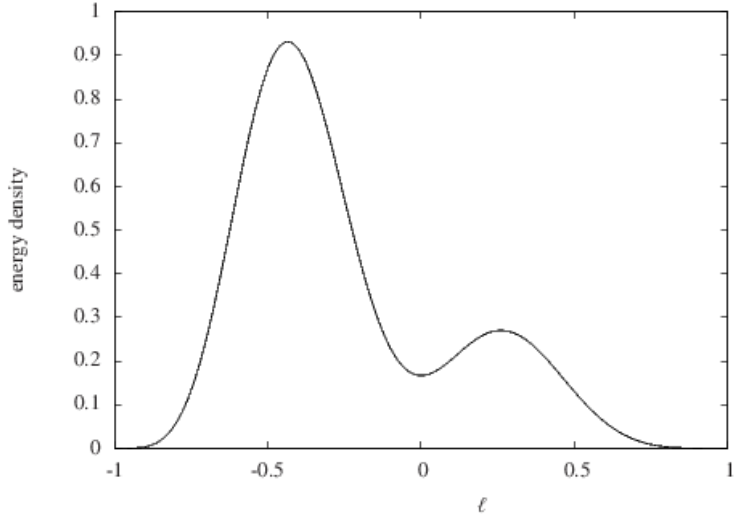}
\caption{The energy density of the tetrahedral 3-monopole along 
a line through the origin and a vertex.}
\label{fig-l3}\end{center}\end{figure}

The energy density can be computed by applying the Laplace--Beltrami
operator, but the result is complicated. It is used to obtain
the energy density isosurface shown in Figure~\ref{fig-1235}.
Along the special line the energy density simplifies to
\be
{\cal E}=\frac{81(1-l^2)^4\bigl(
27(l^8+1)+72l(l^6+1)+1140l^2(l^4+1)+760l^3(l^2+1)+2402l^4\bigr)}
{2(3l^2+4l+3)^4 \, (3l^2-2l+3)^4}\,.
\ee
This expression is plotted in Figure~\ref{fig-l3}.

\subsection{Octahedral 5-monopole}

The JNR ansatz with six poles at the vertices of an octahedron gives a
hyperbolic 5-monopole with octahedral symmetry.
The vertex Klein polynomial of the octahedron is
${\cal O}_v=\xi^5-\xi$, with roots $\infty,\pm 1,\pm i,0$.
The fact that one root is at infinity means that the JNR
ansatz reduces to the 't Hooft ansatz for an instanton \cite{tH,book}, 
where the first term in $\zeta$ is replaced by $1$.

\noindent \ \ The ring of octahedrally invariant homogeneous polynomials is 
generated by 
$o_2=t_2,\,o_4=t_4$ and $o_6=t_3^2$, with $t_2, t_3$ and $t_4$ as in 
(\ref{tetrapoly}). $|\Phi|^2$ can be written in terms of these invariants as
\bea
& &  |\Phi|^2=q_1/q_2 \quad \mbox{where} \nonumber\\
&q_1&=o_2-32o_4+24o_2^2+76o_2^3+192o_6-80o_4o_2-32o_4o_2^2+104o_2^4
-384o_6o_2^2+64o_4^2o_2
\nonumber\\
& &-96o_4o_2^3
 +166o_2^5-32o_4o_2^4
+104o_2^6-80o_4o_2^5+76o_2^7+192o_6o_2^4-32o_4o_2^6+24o_2^8+o_2^9 \,,
\nonumber\\
&q_2&=(3+4o_2-8o_4+10o_2^2+4o_2^3+3o_2^4)^2(1+o_2)^2 \,.
\label{octaHiggs}
\eea
An energy density isosurface is shown in Figure~\ref{fig-1235}.

Along the $X_3$-axis (which passes through two vertices) the above 
simplifies to
\be
|\Phi|^2
={\frac {X_3^{2} \left( X_3^{2}+2X_3-1 \right) ^{2} \left( X_3^{2}-2X_3
-1 \right) ^{2}}{ \left( 3X_3^{4}-2X_3^{2}+3 \right) ^{2} 
\left( X_3^{2}+1 \right) ^{2}}} \,,
\ee
which has zeros at $X_3=0$ and $X_3=\pm(\sqrt{2}-1)$. This is
compatible with there being six zeros of multiplicity $1$ forming 
an octahedron, and an anti-zero at the centre, as in flat space \cite{HS2}.

\begin{figure}[ht]\begin{center}
\includegraphics[width=10cm]{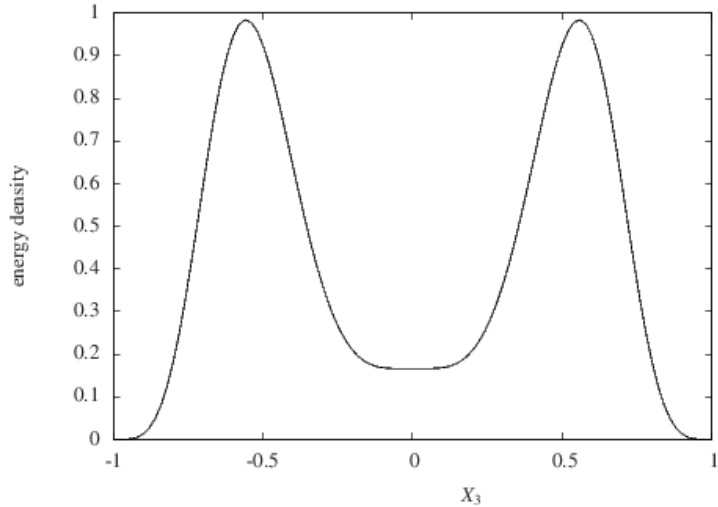}
\caption{The energy density of the octahedral 5-monopole along the $X_3$-axis.}
\label{fig-l5}\end{center}\end{figure}

Along this same axis,
\be
{\cal E}=\frac{(1-X_3^2)^4\bigl(
27(X_3^{16}+1)+120X_3^2(X_3^{12}+1)+5812X_3^4(X_3^8+1)
+14408X_3^6(X_3^4+1)+21474X_3^8
\bigr)}
{2(X_3^2+1)^4 \, (3X_3^4-2X_3^2+3)^4} \,,
\ee
and this is shown graphically in Figure~\ref{fig-l5}.

\subsection{Icosahedral 11-monopole}
The JNR ansatz with twelve poles at the vertices of an icosahedron gives a
hyperbolic 11-monopole with icosahedral symmetry. In an orientation
that has a root at infinity, the vertex Klein polynomial of the icosahedron is
${\cal Y}_v=\xi^{11}+11\xi^6-\xi.$

At charge 11 and higher, it becomes impractical to calculate explicit
expressions for the magnitude of the Higgs field throughout hyperbolic space,
due to the number of terms. However, the Higgs field
of the icosahedral 11-monopole is manageable if restricted to the $X_3$-axis,
and here
\be |\Phi|^2=
{\frac { X_3^{2}\left( 25X_3^{8}+20X_3^{6}-218X_3^
{4}+20X_3^{2}+25 \right) ^{2}}{ \left( 75
 X_3^{10}+55X_3^{8}-2X_3^{6}-2X_3^{4}+
55X_3^{2}+75 \right) ^{2}}}\,.
\ee
Along this axis, which passes through two vertices, there are Higgs 
zeros at $X_3=0$ and $X_3^2=(2\sqrt{17}-1-2\sqrt{11-\sqrt {17}}\,)/5,$ 
compatible with twelve zeros of multiplicity 1 at the vertices of the 
icosahedron and an anti-zero at the centre. 

\section{Circle invariance of ADHM data}\news
The JNR ansatz could be used to construct further hyperbolic
monopoles, mostly with lower symmetry, by having the
poles at more generic positions on the boundary surface of $\bb{H}^3$,
and changing the weights. However, one cannot obtain all hyperbolic
monopoles this way. The dimension of the moduli space of hyperbolic 
monopoles of charge $N$ grows like $4N$, whereas that of the JNR parameter 
space (with poles restricted to a two-dimensional surface) grows like $3N$.
The way to obtain all instantons in $\bR^4$ is to use the ADHM
construction, and in this way one can also obtain all hyperbolic
monopoles. 

In this section we discuss the general class of ADHM data that give rise to
hyperbolic monopoles. That means focussing on circle invariance
first, leaving the possibility of platonic symmetry to later. 
We will use the toroidal coordinate system in $\bR^4$ which
leads to the ball model of $\bb{H}^3$. The ADHM matrices need to
satisfy a number of simultaneous quadratic constraints, and these are
not generally explicitly solvable. 

The ADHM matrices are constant matrices of quaternions, and one also
needs to use the quaternionic representation of a point $x$ in $\bR^4$,
$x=x_4 + x_1 i+ x_2 j + x_3 k$. Then the conformal group of $\bR^4$ 
acts as quaternionic M\"obius transformations
\be
x\mapsto x'=(Ax+B)(Cx+D)^{-1} \,.
\label{genct}
\ee

Platonic ADHM data are symmetric under some finite subgroup $K$ of
$SO(3)$ generated by rotations of the form
\be
\begin{pmatrix} A&B\\ C & D \end{pmatrix}
=\begin{pmatrix} q&0\\ 0 & q \end{pmatrix} \,,
\label{qrotn}
\ee
where $q$ is a unit quaternion representing (in $SU(2)$) an element of 
$K$. The commuting circle action is given by the group of rotations 
\be
\begin{pmatrix} A&B\\ C & D \end{pmatrix}=
\begin{pmatrix} \cos\alphaa&\sin\alphaa\\ -\sin\alphaa & \cos\alphaa
\end{pmatrix} \,.
\label{circle}
\ee
Note that this circle action fixes the 2-sphere given by $x$ a unit 
pure quaternion. This becomes the 2-sphere boundary of $\bb{H}^3$ in 
the ball model.

In terms of the coordinates $X_1,X_2,X_3$ in the ball model,
define the pure quaternion $X=X_1i+X_2j+X_3k$, with $R^2 = |X|^2$. Together
with the coordinate $\chi$ along the circle one obtains the toroidal 
coordinates of $\bR^4$. The corresponding expression for the 
quaternion $x$ is 
\be 
x=\frac{2X+(1-R^2)\sin\chi}{1+R^2+(1-R^2)\cos\chi} \,.
\label{xtoX}
\ee
The circle action (\ref{circle}) corresponds to the rotation 
$\chi\mapsto \chi+\alpha$. 

In standard form, the ADHM data for a charge $N$ instanton are a pair of
quaternionic matrices $L$ and $M$, where $L$ is a row of $N$
quaternions and $M$ is a symmetric $N\times N$ matrix of
quaternions \cite{ADHM}. These are combined into
\be
\hat M = \begin{pmatrix} L\\ M \end{pmatrix}
\label{Mhat}
\ee
and are required to satisfy the quadratic constraints
\be
\hat M^\dagger \, \hat M=R_N \,,
\label{ADHMconstr}
\ee
where $R_N$ is an invertible, real $N\times N$ matrix. The pure quaternion 
part of $\hat M^\dagger \, \hat M$ is required to vanish. From 
$\hat M$ one constructs the ADHM operator 
\be
\Delta(x)=\hat M-Ux=\begin{pmatrix} L\\ M \end{pmatrix}-
\begin{pmatrix} 0\\ 1_N \end{pmatrix}x \,,
\ee
where $1_N$ denotes the $N\times N$ unit matrix. 

Equivalent ADHM data are obtained by applying the transformation
\be
\hat M\mapsto Q\hat M \,, \quad U\mapsto QU \,,
\label{basis}
\ee
where $Q^\dagger Q = 1_{N+1}$, but then the data are (generically) no longer
in standard form. 

We now introduce a stronger set of constraints on the ADHM data than 
(\ref{ADHMconstr}), and show that these are sufficient for the data to be
invariant under the circle action (\ref{circle}). The stronger 
constraints are
\bea
&(i)& M \mbox{\ is pure quaternion and symmetric,}
\label{con1}
\\
&(ii)& \hat M^\dagger \, \hat M=1_N \,,
\label{con2}\\
&(iii)& LM=\mu L \,, \mbox{\ where $\mu$ is a pure quaternion, and $L$
  is non-vanishing.}
\label{con3} 
\eea
We refer to $\mu$ as a left-eigenvalue of $M$.
Properties $(i)$ and $(ii)$ imply that 
\be
L^\dagger L = 1_N + M^2 \,.
\label{Mrelation}
\ee
Another useful relation is 
\be
LL^\dagger=1-|\mu|^2 \,.
\label{Lrelation}
\ee 
To verify this, apply $M$ on the right of $(iii)$ to obtain
$LM^2=\mu LM=\mu^2L=-|\mu|^2L$, where property $(iii)$ has been used
again. Eliminating $M^2$ using (\ref{Mrelation}) gives 
$(LL^\dagger-1+|\mu|^2)L=0$, and the result follows.

Under the general conformal transformation (\ref{genct}) the
ADHM data transform (up to an overall factor on the right) as
\be
U\mapsto U'=UA-\hat MC \,,\ \quad
\hat M\mapsto \hat M'=\hat MD-UB \,,
\label{adhmgenct}
\ee
which are also not in standard form. For the case of the circle 
action (\ref{circle}) the transformation (\ref{adhmgenct}) becomes 
\be
U'=U\cos\alphaa+\hat M\sin\alphaa \,,\ \quad
\hat M'=\hat M \cos\alphaa-U\sin\alphaa \,.
\label{circleMU}
\ee
To show that the constrained ADHM data are circle-invariant we need a matrix 
$Q$ to put these data back into standard form. Using constraints 
$(i)$ to $(iii)$ and the relations (\ref{Mrelation}) and (\ref{Lrelation}),
one finds that the required matrix is
\be
Q=\begin{pmatrix} \cos\alphaa+\mu\sin\alphaa & -L\sin\alphaa \\
L^\dagger \sin\alphaa & 1_N \cos\alphaa -M\sin\alphaa 
\end{pmatrix} \,.
\label{Qalpha}
\ee
It can be checked that $Q^\dagger Q=1_{N+1}$, and 
direct calculation shows that $QU'=U$ and $Q\hat M'=\hat M$, 
so the ADHM data have the required circle invariance, and hence give
rise to a hyperbolic monopole.

To proceed with the ADHM construction of the instanton, and
hence monopole, we need to find an 
$(N+1)$-component column vector $\Psi$ of unit length, 
$\Psi^\dagger\Psi=1$, that solves the linear equation 
\be 
\Psi^\dagger\Delta(x)=0 \,.
\label{adhmlin} 
\ee 
Note that $\Psi$ is unaffected if $\Delta(x)$ is multiplied by a
factor on the right. The instanton gauge potential is then obtained 
from the formula
\be
A_\mu=\Psi^\dagger\partial_\mu \Psi \,, 
\label{adhmpot} 
\ee
where this pure quaternion is regarded as an element of $su(2)$.

Eq.(\ref{xtoX}) shows that when $\chi=0$, then $x=X$.
Hence for circle-invariant data, and setting $\alpha=\chi$, we deduce
that at a point $x$ with toroidal coordinates $X$ and $\chi$,
\be
\Delta(x)=Q^\dagger\Delta(X) \,.
\ee
Here $Q$ is as in (\ref{Qalpha}), with $\alpha=\chi$.
The required vector $\Psi$ can therefore be written in the form
$\Psi=Q^\dagger V$ where $V$ is a unit length column vector, 
$V^\dagger V=1$, that depends only on the pure quaternion $X$ and
solves the linear equation
\be 
V^\dagger\Delta(X)=0 \,.
\label{adhmlinV} 
\ee 
The resulting gauge potential is $\chi$-independent, which is what we
need to interpret the instanton as a hyperbolic monopole. In particular,
the Higgs field of the monopole is
\be
\Phi=A_\chi=V^\dagger \, \Xi \, V \,,
\label{achi}
\ee 
where 
\be
\Xi=Q \, \partial_\chi Q^\dagger=\frac{1}{2}\begin{pmatrix}
-\mu & L\\ -L^\dagger & M \end{pmatrix} \,.
\label{xihalf}
\ee

Interestingly, the left-eigenvalue $\mu$ has a physical meaning, in 
that it is related to the value of the Higgs field at the origin. To 
see this, set $X=0$ and observe that at this point the vector $V$ satisfying 
(\ref{adhmlinV}) is simply
\be
V=\begin{pmatrix}
\mu \\ L^\dagger
\end{pmatrix} \,.
\ee
Substituting this into the expression (\ref{achi}) yields
$\Phi(0)=-\frac{1}{2}\mu$. The gauge invariant quantity is
$|\Phi(0)|=\frac{1}{2}|\mu|$.

As mentioned earlier,
Braam and Austin have previously discussed hyperbolic monopoles in
terms of ADHM data invariant under a circle action \cite{BA}. 
Their analysis is for all half-integer $p$ and their approach used the circle 
action on $\bR^4$ that leads to the upper half space version of 
$\bb{H}^3$. 
Our analysis, adapted to the hyperbolic ball, 
is more restricted and only deals with the case
$p=\half$. Its advantage is that we will be able to explicitly 
find hyperbolic monopoles with platonic symmetry.
To make a connection to the work of Braam and Austin would require
proving a correspondence between ADHM data that satisfies our  
constraints $(i)$ to $(iii)$ and ADHM data with complex entries.
This would also clarify the issue of whether or not our sufficient constraints
are also necessary, which at the moment is unknown. 

\section{Hyperbolic monopoles from ADHM data}\news
In this section we give explicit examples of ADHM data satisfying the 
constraints $(i)$ to $(iii)$ discussed in Section 4, and which therefore 
give rise to hyperbolic monopoles. Many have platonic symmetry. 
Some of the examples reproduce results obtained using the JNR ansatz.

For $N=1$ and $N=2$ we can directly find ADHM data satisfying the
constraints. For larger $N$, we make use of ADHM 
matrices that were previously constructed to give platonically 
symmetric instantons \cite{LM,SiSu,Hou}. These can be written down 
explicitly, after some analysis involving the representation theory of the
relevant platonic symmetry group, $K$. They are seen to satisfy the
constraints provided one fixes their normalisation 
suitably, which corresponds to fixing the scale of the instanton.

It was not recognised previously that these platonic instantons may have
an additional circle invariance, and hence correspond to hyperbolic 
monopoles.  

\subsection{$N=1$}
An admissible $\hat M$, satisfying the constraints, is
\be
\hat M=\begin{pmatrix} 
\sqrt{1-a^2} \\ ai 
\end{pmatrix} \,,
\ee
with $|a|<1$ and $\mu=ai$. This gives a hyperbolic 1-monopole with its centre
along the $X_1$-axis at $X_1=(1-\sqrt{1-a^2})/a$. The Higgs field at
the origin has magnitude $|\Phi(0)|=\frac{1}{2}|\mu|=\frac{1}{2}|a|$.
The squared magnitude of the Higgs field at a general point in the unit ball 
is given by
\be
|\Phi|^2=\frac{4R^2-4aX_1(1+R^2)+a^2(4X_1^2+(1-R^2)^2)}{4(1+R^2-2aX_1)^2}.
\ee
The simplest example is   
\be
\hat M=\begin{pmatrix} 
1 \\ 0
\end{pmatrix} \,,
\ee
with the monopole centred at the origin. The formulae (\ref{1-Higgs}) and
(\ref{1-energy}) for the Higgs field magnitude and energy density are easily
rederived. 

All of these monopoles are spherically symmetric about their centres,
but only in the last case is the symmetry group the standard $SO(3)$
that we have been discussing. 

\subsection{$N=2$}
An axially symmetric 2-monopole, centred at the origin, is
obtained from
\be
\hat M=\frac{1}{2}
\begin{pmatrix} 
\sqrt{2} & \sqrt{2}k\\
i&j\\
j&-i
\end{pmatrix} \,.
\ee
This can be extended to a one-parameter family of non-axially
symmetric monopoles, still centred at the origin. This family
illustrates the $90^\circ$ scattering of $N=2$ monopoles, familiar from
monopoles in $\bR^3$ \cite{AH}. $\hat M$ has the form, satisfying the
constraints, 
\be
\hat M=\frac{1}{2}
\begin{pmatrix} 
\sqrt{2(1-a^2)} & \sqrt{2(1-a^2)}k\\
(1-a)i&(1+a)j\\
(1+a)j&-(1-a)i
\end{pmatrix} \,,
\ee
where $a\in(-1,1)$. The axial case is recovered when $a=0$. 
For this family, the left-eigenvalue of $M$
is $\mu=-ai$, so it vanishes only for the axial example.

$|\Phi|^2$ can be computed at all points in the unit
ball, and is
\bea
& &  |\Phi|^2=q_1/q_2 \quad \mbox{where} \nonumber\\
&q_1&=(a^4+1)\rho^4
-2a^3(X_1^2-X_2^2)((1-R^2)^2+2\rho^2)
+a^2((1-R^4)^2-8R^2\rho^2+6\rho^4-16X_1^2X_2^2)
\nonumber\\
& &+2a(X_1^2-X_2^2)(3R^4+2R^2+3-2\rho^2)
+4R^2(1+R^2)^2-4\rho^2(1+R^4) \,,
\nonumber\\
&q_2&=
\bigl(
-2a^2\rho^2+4a(X_1^2-X_2^2)+2((1+R^2)^2-\rho^2)\bigr)^2 \,,
\eea
and where $\rho^2 = X_1^2 + X_2^2$. Note that when $a=0$ this expression 
reverts to the axial form (\ref{axial2jnr}) obtained earlier using JNR data.
The symmetry under a change of sign of $a$ accompanied by an exchange
of $X_1$ and $X_2$ is clear.

\begin{figure}[ht]\begin{center}
\includegraphics[width=16cm]{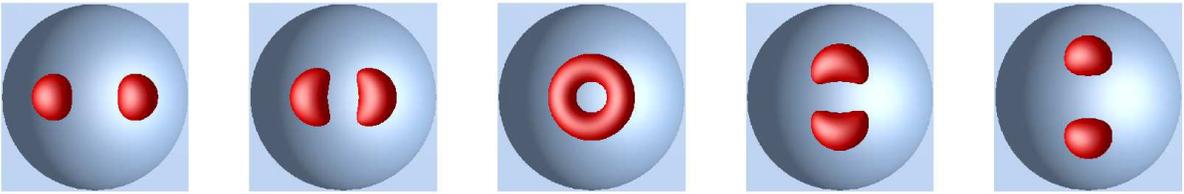}
\caption{Energy density isosurfaces for the 2-monopole with
$a=-0.5, -0.25, 0, 0.25, 0.5.$
The boundary of hyperbolic space is also indicated.
}
\label{fig-2scat}\end{center}\end{figure}

For $a\in(-1,0]$ the two zeros of the Higgs field are on the $X_1$-axis
at the positions
\be
X_1^2=\frac{a^2-3+\sqrt{(1-a^2)(9-a^2)}}{2a} \,.
\ee
For $a\in[0,1)$ the Higgs zeros are on the  $X_2$-axis, as expected 
from the above symmetry under $a\mapsto -a$.
Energy density isosurfaces for several members of this one-parameter
family are displayed in Figure~\ref{fig-2scat}.

\subsection{Tetrahedral $N=3$}
The ADHM data with the normalisation required to satisfy the
constraints are of the form \cite{Hou}
\be
\hat M=\frac{1}{\sqrt{3}}
\begin{pmatrix} 
i&j&k\\
0&k&j\\
k&0&i\\
j&i&0
\end{pmatrix} \,,
\ee
with $\mu = 0$. From this one obtains the Higgs field and energy
density of the tetrahedrally symmetric 3-monopole obtained earlier 
using JNR data.

\subsection{Cubic $N=4$}
\begin{figure}[ht]\begin{center}
\includegraphics[width=7cm]{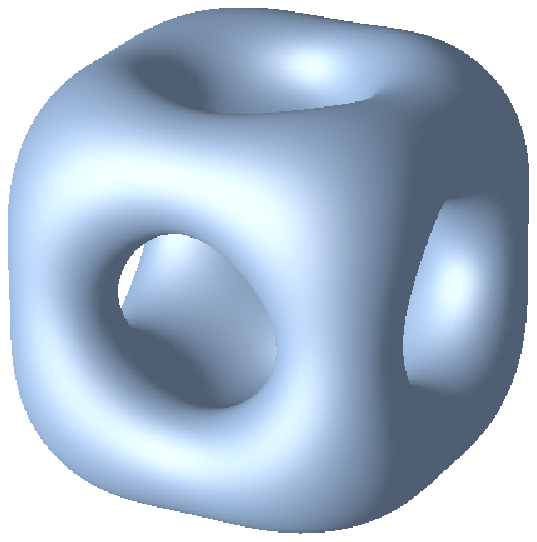}
\includegraphics[width=7cm]{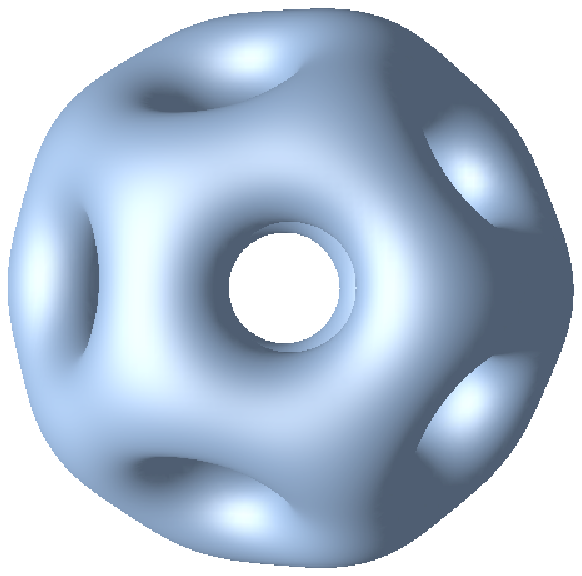}
\caption{Energy density isosurfaces for the cubic 4-monopole
and the dodecahedral 7-monopole.}
\label{fig-e47}\end{center}\end{figure}

This is the first platonic example that cannot be obtained using the
JNR ansatz. Here
\be
\hat M=\frac{1}{2\sqrt{2}}
\begin{pmatrix} 
\sqrt{2}&\sqrt{2}i&\sqrt{2}j&\sqrt{2}k\\
0&-j-k&-k-i&-i-j\\
-j-k&0&j-i&i-k\\
-k-i&j-i&0&k-j\\
-i-j&i-k&k-j&0
\end{pmatrix} \,.
\ee
This is a special case of the ADHM data found in \cite{LM}, with the
normalisation, and hence the instanton scale size, fixed to satisfy
the constraints. The hyperbolic monopole has
\bea
& &  |\Phi|^2=q_1/q_2 \quad \mbox{where} \nonumber\\
&q_1&=
-18o_2^2+54o_4+54o_2^3+108o_6+18o_4o_2+153o_2^4-216o_6o_2+9o_4^2-54o_4o_2^2
+18o_4o_2^3
\nonumber\\
& &+54o_2^5+108o_6o_2^2+54o_4o_2^4-18o_2^6 \,,
\nonumber\\
&q_2&=4(2o_2^4+4o_2^3+3o_2^2+3o_4+4o_2+2)^2 \,,
\eea
with $o_2,o_4$ and $o_6$ the octahedral polynomials as in (\ref{octaHiggs}).
Along the line $X_1=X_2=X_3=l/\sqrt{3}$ (which passes through two
cubic vertices) the above simplifies to
\be
 |\Phi|^2=\frac{4l^6}{(l^2+1)^2(l^4+1)^2} \,.
\ee
This is zero only at the origin, in agreement with the fact that
for the cubically symmetric 4-monopole in $\bR^3$ there are no 
anti-zeros of the Higgs field. Along this line the energy density is
\be
{\cal E}={\frac {{l}^{2} \left( 9\,{l}^{12}+102\,{l}^{10}+283\,{l}^{8}+396
\,{l}^{6}+283\,{l}^{4}+102\,{l}^{2}+9 \right)  \left( 1-l^2 \right) ^{4
}}{ 2\left( {l}^{4}+1 \right) ^{4} \left( {l}^{
2}+1 \right) ^{4}}} \,.
\ee
An energy density isosurface for the cubic 4-monopole is 
presented in Figure~\ref{fig-e47}.

\begin{figure}[ht]\begin{center}
\includegraphics[width=16cm]{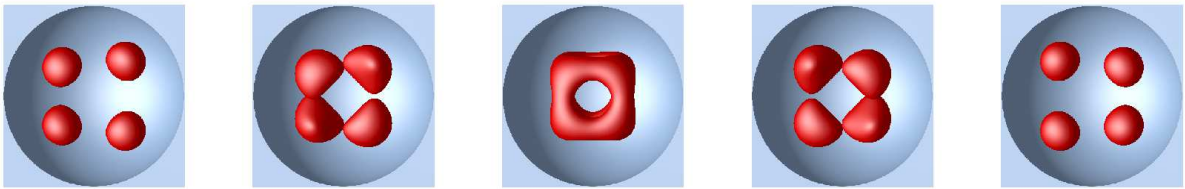}
\caption{Energy density isosurfaces for the 4-monopole with
$a=-0.4 \,, -0.2 \,, 0 \,, 0.2 \,, 0.4 \,.$ The boundary of 
hyperbolic space is also indicated.
}
\label{fig-4scat}\end{center}\end{figure}

The cubically symmetric data can be extended to a one-parameter family
with tetrahedral symmetry, as in \cite{LM}, 
\be
\hat M=
\left( \begin {array}{cccc} b\sqrt {2}&b\sqrt {2}i&b\sqrt {2
}j&b\sqrt {2}k\\ \noalign{\medskip}a \left( i+j+k \right) &-b \left( j
+k \right) &-b \left( k+i \right) &-b \left( i+j \right) 
\\ \noalign{\medskip}-b \left( j+k \right) &a \left( i-j-k \right) &b
 \left( j-i \right) &b \left( i-k \right) \\ \noalign{\medskip}-b
 \left( k+i \right) &b \left( j-i \right) &a \left( -i+j-k \right) &b
 \left( k-j \right) \\ \noalign{\medskip}-b \left( i+j \right) &b
 \left( i-k \right) &b \left( k-j \right) &a \left( -i-j+k \right) 
\end {array} \right) \,,
\ee
where, to satisfy the constraints, $b=\sqrt{(1-3a^2)/8}$ with 
$\sqrt{3}a\in(-1,1)$. 
The cubic case is recovered when $a=0$ and $b=1/2\sqrt{2}$. 
For this tetrahedral family it can be checked that the left-eigenvalue
is $\mu=a(i+j+k)$, so it vanishes only for the cubic example.
Energy density isosurfaces for several members of this one-parameter
family are displayed in Figure~\ref{fig-4scat}.
\subsection{Dodecahedral $N=7$}

The existence of icosahedrally symmetric ADHM data with $N=7$ was
established in \cite{SiSu}. With a suitable normalisation, the data
satisfy the constraints $(i)$ to $(iii)$, with $\mu=0$, and give a
hyperbolic 7-monopole of dodecahedral form. $\hat M$ is
\be
\hat M= \half
\begin{pmatrix}1&i&j&k&0&0 & 0\\
0&0& 0 & 0 & i&j&k\\
0&0&0&0&0&\tau k & \tau\iv j\\
0&0&0&0&\tau\iv k &0&\tau i\\
0&0&0&0&\tau j&\tau\iv i &0 \\
i& 0&\tau\iv k & \tau j & 0 & 0 & 0\\
j& \tau k & 0 &\tau\iv i&0&0&0  \\
k & \tau\iv j & \tau i & 0&0&0&0\\
\end{pmatrix} \,,
\label{adhmY7}
\ee
where $\tau=\frac{1}{2}(\sqrt{5}+1)$. 

The ring of icosahedrally invariant homogeneous polynomials is
generated by three polyomials of degrees two, six and ten,
\bea
y_2&=& X_1^2+X_2^2+X_3^2 \,, \\ 
y_6&=&2\tau(X_1^4X_2^2+X_2^4X_3^2+X_3^4X_1^2)
-2\tau^{-1}(X_1^4X_3^2+X_2^4X_1^2+X_3^4X_2^2)-8X_1^2X_2^2X_3^2 \,,\nonumber \\
y_{10}&=& 
10X_1^{2}X_2^{2}X_3^{2} \left( 5X_1^{2}X_2^{2}+5X_1^{2}X_3^{2}+5X_2^{2}X_3^{2}
-2X_1^{4}-2X_2^{4}-2X_3^{4} \right) \nonumber \\
& &+2\sqrt{5} \bigl( X_1^{2}X_2^{8}-X_2^{2}X_1^{8}+X_1^{8}X_3^{2}
-X_3^{8}X_1^{2}+X_3^{8}X_2^{2}-X_2^{8}X_3^{2}
 \nonumber \\
& &-{\tau}\left( X_1^{4}X_2^{6}+X_2^{4}X_3^{6}+X_3^{4}X_1^{6} \right) 
-{\tau^{-1}} \left( X_1^{4}X_3^{6}+X_2^{4}X_1^{6}+X_3^{4}X_2^{6}
\right) \bigr) \,. 
\nonumber
\eea
The squared magnitude of the Higgs field can be expressed in terms of 
these, and is
\bea
& &  |\Phi|^2=q_1/q_2 \quad \mbox{where} \nonumber\\
&q_1&=8y_2^3+60y_6+56y_2^4+180y_6y_2+194y_2^5+185y_6y_2^2+25y_{10}
+392y_2^6-20y_6y_2^3
\nonumber\\& &
+500y_2^7-50y_{10}y_2^2+50y_6^2y_2-210y_6y_2^4+392y_2^8-20y_6y_2^5
+194y_2^9+25y_{10}y_2^4
\nonumber\\& &
+185y_6y_2^6+56y_2^{10}+180y_6y_2^7+8y_2^{11}+60y_6y_2^8 \,,
\nonumber\\
&q_2&=
2(1+y_2)^2(2+4y_2+6y_2^2+6y_2^3+5y_6+6y_2^4+4y_2^5+2y_2^6)^2 \,. 
\eea
An energy density isosurface for the dodecahedral 
7-monopole is presented in Figure~\ref{fig-e47}.

\subsection{Icosahedral $N=17$}
Icosahedrally symmetric ADHM data for an instanton with $N=17$ are 
given in \cite{Su17}. In this example the matrix $\hat M$ is quite 
large, so we do not reproduce it here. However it can be checked that 
it does indeed satisfy the constraints $(i)$ to $(iii)$ for circle 
invariance, after multiplication by a scale factor of $\frac{1}{2}$ 
compared to the normalisation presented in \cite{Su17}. The 
left-eigenvalue $\mu$ again vanishes. Although we have not attempted 
to compute the Higgs field and energy density of the resulting
17-monopole, the known properties of the instanton make it clear that 
the polyhedron associated with this example is
the truncated icosahedron, familiar as the buckyball.

\section{Rational maps}\news
One of the achievements of earlier work on hyperbolic monopoles
\cite{At,BA} was the establishment of a one-to-one correspondence between 
charge $N$ monopoles and rational maps (from the Riemann sphere to itself) 
of degree $N$. 

We have not succeeded in constructing a rational map from 
a general hyperbolic monopole, in our formalism. However in the
cases where $\mu = 0$ we have a good candidate. This appears to be an 
analogue of the Jarvis map for Euclidean monopoles \cite{Jar}. In particular, 
the map has platonic symmetry if the monopole has platonic symmetry.

To proceed, let $X$ be a unit pure quaternion. This can be
identified with a point on the Riemann sphere with complex 
coordinate $\xi$ by writing
\be
X=\frac{1}{1+|\xi|^2}\bigg(2 \, {\rm Re}(\xi)i
+2 \, {\rm Im}(\xi)j +(|\xi|^2-1)k \bigg) \,.
\ee
Next, use the ADHM data to define the quaternion function of $X$,
\be
f(X)=L(M-X)^{-1}L^\dagger \,,
\label{fratmap}
\ee
where $M-X$ means $M - X \, 1_N$. Using constraint $(i)$, it is easy 
to see that $f$ is a pure quaternion for all $X$. 

If $\mu=0$, then $|f|^2 = 1$ so $f$ is a unit pure quaternion. To prove this, 
note first that as $f$ is a pure quaternion,
\be
|f|^2 = -L(M-X)^{-1}L^\dagger L(M-X)^{-1}L^\dagger
\ee
and therefore, using (\ref{Mrelation}),
\be
|f|^2 = -L(M-X)^{-1} (1_N + M^2)(M-X)^{-1}L^\dagger \,.  
\label{modf}
\ee
Next, as $X$ is a unit pure quaternion, $(M-X)^{-1} 
= (-X(1_N + XM))^{-1} = (1_N + XM)^{-1}X$, from which follows the 
formal series expansion
\be
(M-X)^{-1} = X - XMX + XMXMX - XMXMXMX + \cdots \,.
\label{MXseries}
\ee
Inserting this twice in (\ref{modf}), using $X^2 = -1$, and collecting 
terms, we obtain the relatively simple series
\be
|f|^2 = -L(-1_N + MX + XM - MXMX - XMXM + MXMXMX + XMXMXM - \cdots)
L^\dagger \,.
\ee
If $\mu = 0$, then $LM = 0$, using constraint $(iii)$, and by conjugation 
$ML^\dagger = 0$, so all terms in this series except the first vanish, 
and by (\ref{Lrelation}), $L L^\dagger = 1$, so $|f|^2 = 1$, as claimed.  

The unit pure quaternion $f$, like $X$, may be identified with 
a Riemann sphere coordinate ${\cal R}$ via
\be
f=\frac{1}{1+|{\cal R}|^2}\bigg(2 \, {\rm Re}({\cal R})i
+2 \, {\rm Im}({\cal R})j +(|{\cal R}|^2-1)k \bigg) \,.
\ee
The function $f(X)$ therefore gives a function ${\cal R}(\xi)$, our
candidate rational map, although at this point ${\cal R}$ could also 
depend on $\bar \xi$. 

To prove that ${\cal R}(\xi)$ 
is holomorphic, we recall the complex structure on the Riemann sphere 
of unit pure quaternions. Let $X+\delta X$ be a unit pure quaternion 
infinitesimally separated from $X$. Since $X^2 = -1$ and 
$(X+\delta X)^2 = -1$, to linear order $X\delta X + \delta X X = 0$. 
This is the condition for $\delta X$ to be tangent to the Riemann sphere 
at $X$. The complex structure operation at $X$ is right multiplication by
$X$, $\delta X \to \delta X X$, which acts on the tangent space and 
whose square is multiplication by $-1$. Similarly, the complex
structure on the target Riemann sphere is right multiplication by $f$.

Now consider $\delta f$, effectively the derivative of $f(X)$, defined 
as $f(X+\delta X) - f(X)$ truncated at linear order in $\delta X$. 
From (\ref{fratmap}), we find
\be
\delta f = L(M-X)^{-1} \delta X (M-X)^{-1} L^\dagger \,.
\ee
${\cal R}(\xi)$ will be holomorphic if the effect of replacing $\delta X$ 
by $\delta X X$ is to replace $\delta f$ by $\delta f f$. So we need 
to show that
\be
L(M-X)^{-1} \delta X X (M-X)^{-1} L^\dagger
= L(M-X)^{-1} \delta X (M-X)^{-1} L^\dagger L(M-X)^{-1}L^\dagger \,.
\ee
The first few factors on each side are the same, so it is sufficient
to show that 
\be
X (M-X)^{-1} L^\dagger = (M-X)^{-1} L^\dagger L(M-X)^{-1}L^\dagger \,.
\ee
Using the series (\ref{MXseries}) again, and the relations
(\ref{Mrelation}) and $X^2 = -1$, this reduces to
\be
(-1_N + MX - MXMX + \cdots)L^\dagger
= (-1_N + MX + XM - MXMX - XMXM + \cdots)L^\dagger \,.
\ee
As $ML^\dagger = 0$, the left and right hand sides are equal, so  
${\cal R}(\xi)$ is indeed holomorphic.

In fact, for all our platonic monopole examples with $\mu=0$, we find 
that ${\cal R}$ is a rational function of $\xi$ whose 
degree equals the monopole charge. In detail, for the spherical 
1-monopole and the axial 2-monopole the maps are ${\cal R}=\xi$ and 
${\cal R}=\xi^2$, respectively. More complicated examples are provided 
by the tetrahedral 3-monopole and the dodecahedral 7-monopole, where 
the above construction yields the rational maps
\be
{\cal R}=\frac{\sqrt{3}i\xi^2-1}{\xi^3-\sqrt{3}i\xi}
\quad\quad \mbox{and} \quad\quad
{\cal R}=\frac{7\xi^6-7\sqrt{5}\xi^4-7\xi^2-\sqrt{5}}{\sqrt{5}\xi^7
+7\xi^5+7\sqrt{5}\xi^3-7\xi} \,.
\ee
These maps agree with the Jarvis maps presented in \cite{HMS}
for the corresponding platonic monopoles in Euclidean space.

\section{Beyond $p=\half$}\news
The JNR construction of hyperbolic monopoles, presented in Section 3, applies
only to the case $p=\half.$ To obtain hyperbolic monopoles with other
half-integer values of $p$ requires a placement of the poles out of the 
plane $r=0.$ However, once the poles do not lie in this fixed set of the 
circle action then it is more difficult to arrange for circle invariance.
In fact the poles must all have the same weight and be equally spaced on 
a circle. This arrangement was first identified by Nash \cite{Na}, who
noted that the 1-monopole is obtained from a JNR instanton of
charge $2p$ by placing equal weight poles at the vertices of a regular 
$(2p+1)$-gon inscribed in the circle $r=1$ with $x_1=x_2=0.$
Note that the case $p=\half$ is again special here, in that this description
of the 1-monopole involving two poles is equivalent to the earlier 
description, where two poles are placed in the $r=0$ plane. This dual 
description follows from the $SO(4)$ symmetry of the 1-instanton. 

For $p\ne \half$ the above polygonal arrangement is the only option
for circle invariance, so more complicated monopoles with platonic
symmetry are beyond this JNR ansatz. The ADHM construction, or its 
circle-invariant formulation by Braam and Austin, is then required. 
The possibility of an explicit construction of platonic
monopoles is therefore not guaranteed, and certainly requires a full
description of the non-standard action of $SO(3)$ on such data.

We now return to the discussion of circle-invariant ADHM data, as
described in Section 4, and make some comments regarding its extension beyond
$p=\half.$ First of all, it is not apparent from the constraints 
$(i)$ to $(iii)$, that these apply to hyperbolic monopoles
with $p=\half.$ This is only evident upon examination of the 
associated compensating matrix (\ref{Qalpha}), which reveals that
it involves only the half-angle $\alpha/2,$ hence $p=\half.$
To illustrate the challenges that arise in attempting to go 
beyond $p=\half$
within this formalism, 
we present the construction of the 1-monopole with $p=1.$ 

Consider the 2-instanton given by the following ADHM data,
\be
\hat M = 
\frac{1}{\sqrt{3}}\begin{pmatrix} 2\sqrt{2} & 0 \\
0 & 1\\
1 & 0 
\end{pmatrix}.
\label{p1data}
\ee
$\hat M$ is obviously $SO(3)$ symmetric, as it contains only real
entries and hence is invariant under rotations given by (\ref{qrotn})
for all unit quaternions $q.$
To show that this data is also circle-invariant,
we need to demonstrate that the circle action (\ref{circleMU})
can be compensated. This requires that 
\be 
QU'P=U \quad\quad \mbox{and} \quad\quad Q\hat M' P=\hat M,
\ee 
for some $3\times 3$ matrix $Q$ 
with $Q^\dagger Q=1_3,$ and some invertible $2\times 2$ matrix $P.$
In the case that $p=\half,$ the corresponding transformation (\ref{basis}), 
mapping to equivalent ADHM data, did not explicitly include the
matrix $P$ because it simplifies to the identity matrix.
For $p\ne \half$ this simplification is no longer possible.

It may be verified that the required matrices are given by
\be
Q=
\begin{pmatrix} 
\frac{1}{3}(1+2\cos\alpha) & 
-\sqrt{\frac{2}{3}}\sin\alpha & 
\frac{\sqrt{2}}{3}(1-\cos\alpha)
\\
\sqrt{\frac{2}{3}}\sin\alpha & 
\cos\alpha &
-\frac{1}{\sqrt{3}}\sin\alpha 
\\
\frac{\sqrt{2}}{3}(1-\cos\alpha) &
\frac{1}{\sqrt{3}}\sin\alpha &
\frac{1}{3}(2+\cos\alpha) 
\end{pmatrix}
\quad\mbox{and}\quad
P=\begin{pmatrix} 
\cos\alphaa & 
\frac{1}{\sqrt{3}}\sin\alphaa \\
-\sqrt{3}\sin\alphaa &
\cos\alphaa
\end{pmatrix}.
\ee
As before, the value of $p$ is not apparent in the ADHM data (\ref{p1data}),
but is evident in the compensating matrix $Q,$ which involves functions of the 
angle $p\alpha$ with $p=1.$ This contrasts with the
$p=\half$ compensating matrix (\ref{Qalpha}), 
which involves functions of the half-angle $\alphaa.$

The ADHM construction of the hyperbolic monopole proceeds as
before by setting $\alpha=\chi$ so that
\be
\Phi=A_\chi=V^\dagger \, \Xi \, V \,,
\ee 
where 
\be
\Xi=Q \, \partial_\chi Q^\dagger=\frac{1}{\sqrt{3}}\begin{pmatrix}
0 & \sqrt{2} & 0\\
-\sqrt{2} & 0 & 1 \\
0 & -1 & 0
\end{pmatrix}
\ee
and $V$ is a unit length solution of (\ref{adhmlinV}). $\Xi$ has a 
different form to the $p=\half$ expression (\ref{xihalf}). 
The corresponding magnitude of the Higgs field reproduces 
the $p=1$ spherically symmetric 1-monopole formula (\ref{pone}).

The $p=1$ ADHM data (\ref{p1data}) does not satisfy any of our three
constraints $(i)$ to $(iii)$ applicable to $p=\half$ monopoles.
For example,
\be
\hat M^\dagger \hat M = 
\begin{pmatrix} 3 & 0\\ 0 & \frac{1}{3} \end{pmatrix}.
\ee
This demonstrates that our new constraints are specific to 
$p=\half$ monopoles, and it is not known whether there is an
appropriate generalisation to other half-integer values of $p.$

\section{Conclusion}\news
We have described methods to construct explicit examples
of hyperbolic monopoles using circle-invariant instantons, both
within the JNR and ADHM approaches. Several examples have been presented
in detail, including a number with platonic symmetry.  
These solutions provide analytic information about hyperbolic monopoles that
complements similar (though sometimes numerical)
 results known for monopoles in flat space.

The platonic examples discussed in this paper are singled out by their
symmetry properties, but energetically they are simply points in a large 
moduli space of BPS hyperbolic monopoles. However, an additional motivation to 
study these symmetric examples is provided by the related problem for
monopoles in four-dimensional Anti-de Sitter spacetime, which has 
three-dimensional hyperbolic space as its constant time slices.
In the Anti-de Sitter case there is essentially a unique minimal energy
monopole for each charge, rather than a moduli space, and numerical
results \cite{Su3} suggest that the minimal energy monopole is often of
the symmetric type considered here. Analytic formulae in the hyperbolic
case may therefore be useful in understanding the features of
monopoles observed in the less tractable Anti-de Sitter situation. 
 
There are many similarities between monopoles and Skyrmions \cite{book},
and this work provides the opportunity to investigate a new connection.
It has been observed \cite{AS} that Skyrmions with massive pions may be
approximated by Skyrmions with massless pions in hyperbolic space, which
in turn can be approximated by the holonomy along circles of suitable 
instantons. If the instantons are taken to be circle-invariant, as in
the present paper, then the computation of the holonomy involves no
integration, and the result is an approximation of Skyrmions by the 
exponential of the Higgs field of a hyperbolic monopole, in a suitable 
gauge. The explicit hyperbolic monopoles presented in this paper will allow a 
detailed investigation of this issue.    
 
\section*{Acknowledgements}
We thank Sir Michael Atiyah for stimulating our interest in this project
and Michael Singer for useful discussions.
We acknowledge EPSRC and STFC for grant support.


\begin{thebibliography}{99}
 
\bibitem{Wa8} R.~S. Ward, 
Two Yang--Mills--Higgs monopoles close together,
\textit{Phys. Lett.} \textbf{B102}, 136 (1981).

\bibitem{FHP} P. Forg\'acs, Z. Horv\'ath and L. Palla,
Exact multimonopole solutions in the Bogomolny--Prasad--Sommerfield limit,
\textit{Phys. Lett.} \textbf{B99}, 232 (1981); 
Solution-generating technique for self-dual monopoles,
\textit{Nucl. Phys.} \textbf{B229}, 77 (1983).

\bibitem{PR} M.~K. Prasad and P. Rossi,
Construction of exact Yang--Mills--Higgs multimonopoles of arbitrary charge,
\textit{Phys. Rev. Lett.} \textbf{46}, 806 (1981).

\bibitem{Nahm} W. Nahm, {The construction of all self-dual
multimonopoles by the ADHM method}, in \textit{Monopoles in Quantum Field
Theory}, eds. N.~S. Craigie, P. Goddard and W. Nahm, Singapore,
World Scientific, 1982.

\bibitem{Hit} N.~J. Hitchin, 
On the construction of monopoles,
\textit{Commun. Math. Phys.} \textbf{89}, 145 (1983).

\bibitem{HMM} N.~J. Hitchin, N.~S. Manton and M.~K. Murray,
Symmetric monopoles,
\textit{Nonlinearity} \textbf{8}, 661 (1995).

\bibitem{HS2} C.~J. Houghton and P.~M. Sutcliffe, 
Octahedral and dodecahedral monopoles,
\textit{Nonlinearity} \textbf{9}, 385 (1996).

\bibitem{HS1} C.~J. Houghton and P.~M. Sutcliffe,
Tetrahedral and cubic monopoles,
\textit{Commun. Math. Phys.} \textbf{180}, 343 (1996).

\bibitem{Su4} P.~M. Sutcliffe,
Monopole zeros,
\textit{Phys. Lett.} \textbf{B376}, 103 (1996).

\bibitem{At}
M.~F. Atiyah, {Magnetic monopoles in hyperbolic spaces},
in \textit{M. Atiyah: Collected Works, vol. 5},
Oxford, Clarendon Press, 1988.

\bibitem{Su3} P.~M. Sutcliffe,
Monopoles in AdS,
\textit{JHEP} \textbf{1108}, 032 (2011).

\bibitem{MSam} N.~S. Manton and T.~M. Samols,  
Skyrmions on $S^3$ and $H^3$ from instantons, 
\textit{J. Phys.} \textbf{A23}, 3749 (1990).

\bibitem{AS} M.~F. Atiyah and P.~M. Sutcliffe, 
Skyrmions, instantons, mass and curvature,
\textit{Phys. Lett.} \textbf{B605}, 106 (2005).

\bibitem{JNR} R. Jackiw, C. Nohl and C. Rebbi, 
Conformal properties of pseudoparticle configurations,
\textit{Phys. Rev.} \textbf{D15}, 1642 (1977).

\bibitem{CF} E. Corrigan and D.~B. Fairlie,
Scalar field theory and exact solutions to a classical $SU(2)$
gauge theory,
\textit{Phys. Lett.} \textbf{B67}, 69 (1977).

\bibitem{ADHM} M.~F. Atiyah, N.~J. Hitchin, V.~G. Drinfeld and Yu.~I. Manin,
Construction of instantons,
\textit{Phys. Lett.} \textbf{A65}, 185 (1978).

\bibitem{JT} A. Jaffe and C. Taubes, \textit{Vortices and
Monopoles}, Boston, Birkh\"auser, 1980.

\bibitem{Cha} A. Chakrabarti, 
Construction of hyperbolic monopoles,
\textit{J. Math. Phys.} \textbf{27}, 340 (1986).

\bibitem{Na} C. Nash, 
Geometry of hyperbolic monopoles,
\textit{J. Math. Phys.} \textbf{27}, 2160 (1986).

\bibitem{AH} M.~F. Atiyah and N.~J. Hitchin,
\textit{The Geometry and Dynamics of Magnetic Monopoles},
Princeton University Press, 1988.

\bibitem{book} N. Manton and P. Sutcliffe,
\textit{Topological Solitons},
Cambridge University Press, 2004.

\bibitem{Shn} Ya. Shnir,
\textit{Magnetic Monopoles},
Berlin Heidelberg, Springer, 2005.

\bibitem{PS} M.~K. Prasad and C.~M. Sommerfield,
Exact classical solution for the 't Hooft monopole and the Julia--Zee dyon,
\textit{Phys. Rev. Lett.} \textbf{35}, 760 (1975).

\bibitem{Bo} E.~B. Bogomolny,
The stability of classical solutions, 
\textit{Sov. J. Nucl. Phys.} \textbf{24}, 449 (1976).

\bibitem{LM} R.~A. Leese and N.~S. Manton,
Stable instanton-generated Skyrme fields with baryon numbers 
three and four, 
\textit{Nucl. Phys.} \textbf{A572}, 575 (1994).

\bibitem{SiSu} M.~A. Singer and P.~M. Sutcliffe,
Symmetric instantons and Skyrme fields,
\textit{Nonlinearity} \textbf{12}, 987 (1999).

\bibitem{BA} P.~J. Braam and D.~M. Austin,
Boundary values of hyperbolic monopoles,
\textit{Nonlinearity} \textbf{3}, 809 (1990).

\bibitem{Jar} S. Jarvis, 
A rational map for Euclidean monopoles via radial scattering,
\textit{J. reine angew. Math.} \textbf{524}, 17 (2000).

\bibitem{MuSi} M.~K. Murray and  M.~A. Singer, 
Spectral curves of non-integral hyperbolic monopoles,
\textit{Nonlinearity} \textbf{9}, 973 (1996); 
On the complete integrability of the discrete Nahm equations,
\textit{Commun. Math. Phys.} \textbf{210}, 497 (2000).

\bibitem{NoRo} P. Norbury and N. Rom\~ao,
Spectral curves and the mass of hyperbolic monopoles,
\textit{Commun. Math. Phys.} \textbf{270}, 295 (2007).

\bibitem{Kle} F. Klein, 
\textit{Lectures on the Icosahedron},
London, Kegan Paul, 1913.

\bibitem{Wa1} R.~S. Ward, 
A Yang--Mills--Higgs monopole of charge 2,
\textit{Commun. Math. Phys.} \textbf{79}, 317 (1981).

\bibitem{tH} G. 't Hooft, unpublished.

\bibitem{Hou} C.~J. Houghton, 
Instanton vibrations of the 3-Skyrmion,
\textit{Phys. Rev.} \textbf{D60}, 105003 (1999).

\bibitem{Su17} P.~M. Sutcliffe,
Instantons and the buckyball,
\textit{Proc. R. Soc. Lond.} \textbf{A460}, 2903 (2004).

\bibitem{HMS} C.~J. Houghton, N.~S. Manton and P.~M. Sutcliffe,
Rational maps, monopoles and Skyrmions,
\textit{Nucl. Phys.} \textbf{B510}, 507 (1998).



\end{thebibliography}
\end{document}